\documentclass{jfm}
\usepackage{graphicx}
\usepackage{epstopdf, epsfig}
\usepackage{color}
\usepackage{siunitx}
\usepackage{url}

\graphicspath{{./Figures/}} 

\shorttitle{Wake of super-hydrophobic falling spheres}
\shortauthor{M. Castagna, N. Mazellier and A. Kourta}

\title{Wake of super-hydrophobic falling spheres: influence of the air layer deformation}

\author{Marco Castagna\aff{1},
  Nicolas Mazellier\aff{1} 
  \corresp{\email{nicolas.mazellier@univ-orleans.fr}}
 \and Azeddine Kourta\aff{1}}

\affiliation{\aff{1}University of Orl\'eans, INSA-CVL, PRISME, EA 4229, 45072, Orl\'eans, France}

\begin{document}

\maketitle

\begin{abstract}
We report an experimental investigation of the wake of free falling super-hydrophobic spheres. The mutual interaction between the air layer (plastron) encapsulating the super-hydrophobic spheres and the flow is emphasised by studying the hydrodynamic performances. It is found that the air plastron adapts its shape to the flow-induced stresses which compete with the surface tension. This competition is characterised by introducing the Weber number $\mathcal{W}e$, whilst the plastron deformation is estimated via the aspect ratio $\chi$. While noticeable distortions are locally observed, the plastron becomes more and more spherical in average (i.e. $\chi \rightarrow 1$) as far as $\mathcal{W}e$ increases. The study of the falling motion reveals that the plastron compliance has a sizeable influence on the wake development. Investigating the lift force experienced by the super-hydrophobic spheres, the onset of wake instabilities is found to be triggered earlier than for smooth spheres used as reference. Surprisingly, it is also observed that the early promotion of the wake instabilities is even more pronounced beyond a critical Weber number, $\mathcal{W}e_c$, which corresponds to a critical aspect ratio $\chi_{c}$. Furthermore,  the magnitude of the hydrodynamic loads is found to be dependent on the average deformation of the gas/liquid interface. Indeed, in comparison to the reference spheres, high deformation achieved for $\chi > \chi_c$ (oblate shape) leads to lift and drag increase, whereas low deformation obtained for $\chi < \chi_c$ (spherical shape) yields lift and drag mitigation. Accordingly, taking into account the plastron deformation provides an attractive way to explain the somehow discordant results reported in other studies at comparable Reynolds numbers. These results suggest that the amount of vorticity produced at the body surface and then released in the wake is strongly impacted by the plastron compliance. If confirmed by additional studies, our findings would imply that plastron compliance and its feedback on the flow, which are currently neglected in most theoretical works and numerical simulations, must be accounted for to design super-hydrophobic surfaces and/or predict their performances.   
\end{abstract}

\section{Introduction}

Among the large variety of materials fashioned for industrial needs, super-hydrophobic (SH) surfaces have attracted an increasing attention since the $90'$s \citep{Quere2008}. Resulting from the combination of surface texturing and chemical repellency \citep{Shirtcliffe2010}, SH surfaces in the so-called Cassie-Baxter state \citep{Cassie1944} are capable to entrap a gas layer, referred to as plastron, in their roughness restricting thereby the solid/liquid contact area. This feature yields slippage, which may have a strong impact in a wide range of engineering applications where wettability control is essential \citep{Zhang2012}. In the framework of drag control, a number of studies have evidenced the beneficial effects of SH surfaces in reducing skin friction at laboratory scale \citep{Rothstein2010}. However, the extrapolation of these results under broader operating conditions is still to be demonstrated \citep{Samaha2012}. One of the main arising issues is the stability of the gas/liquid interface leading eventually to plastron failure (the so-called transition from the Cassie-Baxter state to the Wenzel state), which is a cornerstone to warrant the effectiveness and the sustainability of SH surfaces \citep{Piao2015}. In particular, the effects of flow-induced perturbations on the interface stability and the possible feedback of plastron compliance on the flow still represent a tough challenge for current theoretical studies and numerical simulations.

The range of scales featuring the problem is so huge that a full resolution is unrealistic. Furthermore, solving a two-phase flow involving stiff jumps of density at the gas/liquid interface is a challenging issue. To bypass these constraints, the interface between the gas and the liquid can be modelled by a non-deformable surface neglecting thereby the effect of surface tension. For instance, the slip length model is a popular boundary condition used in a large number of numerical simulations or theoretical studies to mimic SH surfaces from partial slip \citep[see e.g.][]{Min2004, Ybert2007} to perfect slip \citep[see e.g.][]{Lauga2003, Martell2009}. In this approach, the liquid phase flows over patches of slip/no-slip areas arranged either regularly or randomly. Recently, more realistic approaches have been deployed by simulating the flows within the two phases \citep{Gruncell2013, Jung2016} but still avoiding any deformation of the gas/liquid interface. Neglecting the plastron compliance relates to two major assumptions: \emph{i.} there is no possible feedback on the flow due to the shape modification of the boundary condition and \emph{ii.} the plastron is presumed resilient even at high speed flows. The relevance of these assumptions is disputed by experimental observations at both low and high Reynolds numbers. \citet{Byun2008} investigated the flow within microchannels with transverse grooves. They observed curved gas/liquid interface, which induced wavy flow on the grooves for large pitch-to-width ratio. In some cases, plastron depletion was reported. \citet{Kim2017} studied the influence of the gas/liquid interface on the laminar flow over an array of SH pillars spanning the entire height of a microchannel. Both pressure and viscous drag were found to be sensitive to the curvature of the air bubbles entrapped within the pillar cavity. Studying respectively turbulent flows over a flat plate and circular cylinders treated with SH coatings, \citet{Aljallis2013} and \citet{Kim2015} reported performance losses once the air plastron was washed out. This issue has been addressed recently in studies aiming at identifying the limits at which plastron failure is likely to occur. \citet{Piao2015} investigated theoretically the longevity of the plastron in a single 2D groove subjected to harmonic pressure fluctuations and gas diffusivity. They found that for underwater applications where immersion depth is typically \textit{O}$(\mbox{m})$, pressure-induced perturbation is the leading order effect. Their findings revealed that the interface stays stable only for small enough width of the groove, i.e. \textit{O}$(\mbox{\SI{}{\micro\metre}})$. Based on the Direct Numerical Simulations (DNS) of a turbulent channel flow, \citet{Seo2015} investigated the pressure induced by the footprint of the surface texture modelled as a local shear-free condition. The resulting averaged pressure distribution was then used to predict the typical interface deformation over a groove. Although their approach was restricted to an one-way coupling, these authors delineated critical texture size beyond which plastron failure may occur, which was found to depend on both the pattern width and the Weber number expressed in wall units.

This has motivated us to investigate the possible interplay between the flow and the gas/liquid interface, which is at the core of this study. To this end, we designed an experimental set-up to study the hydrodynamic performances of free falling SH spheres, which have been chosen as a prototype geometry. A large number of experimental, numerical and theoretical works have been devoted to the investigation of the wake of free falling/rising bluff bodies \citep[see][and references therein]{Ern2012}. All in all, these studies emphasised the key role of the vorticity generated at the surface of the bodies in the development of the wake \citep[see e.g.][]{Leal1989, Magnaudet2007}. Indeed, the amount of vorticity produced within the boundary layer developing on the surface of the body is at the root of the onset of wake instabilities and has a direct impact on the hydrodynamic forces \citep{Wu2007}. In addition to parameters such as Reynolds number or buoyancy effect, several works pointed out the sizeable influence of both body shape and slip condition at the body surface on the production of vorticity. For freely rising elongated bodies, \citet{Fernandes2007} investigated the critical Reynolds number at which the onset of wake instabilities appears, with respect to the aspect ratio. They reported lower critical Reynolds numbers than those of spheres. Earlier transition was also observed by \citet{Jenny2004} for imperfect spheres produced either by damaging the surface or by displacing the center of gravity. Using 2D DNS at low Reynolds number, \citet{Legendre2009} investigated the effect of slip boundary condition on the production of vorticity at the surface of a circular cylinder. They showed that the slip condition delays the appearance of wake instabilities and yields drag and lift mitigation. Based on a two-phase approach, \citet{Gruncell2013} simulated the steady laminar flow around a sphere encapsulated within an air layer. A significant drag reduction was obtained with respect to the dimensionless plastron thickness. The influence of the shape of bodies with perfect slip was studied by \citet{Mougin2002b} and \citet{Magnaudet2007} who carried out DNS of flows around non-deformable oblate spheroidal bubbles. They evidenced that the maximum surface vorticity increases with the aspect ratio of the bubble. Exploring a wide range of control parameters, they observed that wake instabilities appear for high enough aspect ratio and coincide with the increase of flow-induced loads.

Very few experimental works have been dedicated to the study of the drag of spheres covered by a SH coating. \citet{McHale2009} performed falling experiments of SH acrylic spheres over a wide range of terminal Reynolds number $\Rey_\infty$ (based on the diameter of the sphere and its falling terminal velocity) using several surface treatments. An intermediate sand layer was deposited to vary the surface roughness. The authors investigated the change in terminal velocity and accordingly in drag with and without the presence of the air plastron. More recently, another falling sphere experiment was carried out by \citet{Ahmmed2016} using laser-textured PTFE spheres at both low and high $\Rey_\infty$ using glycerine/water mixtures. Although the ranges of $\Rey_\infty$ investigated by \citet{McHale2009} and \citet{Ahmmed2016} do not overlap, some data were collected at comparable $\Rey_\infty$. Within the Cassie-Baxter state, significant drag reduction ($\approx$ 10\%) was achieved by \citet{McHale2009} for $\Rey_\infty \approx 10^4$. On the contrary, at $\Rey_\infty \approx 5 \times 10^3$, \citet{Ahmmed2016} evidenced a decrease of the terminal velocity by up to 10\% yielding an increase by 20\% of the drag in comparison to standard spheres. These somehow discordant results illustrate that a comprehensive understanding of the flows over SH bluff bodies is still missing.

\begin{figure}
  \centerline{\includegraphics[width = 0.95 \textwidth]{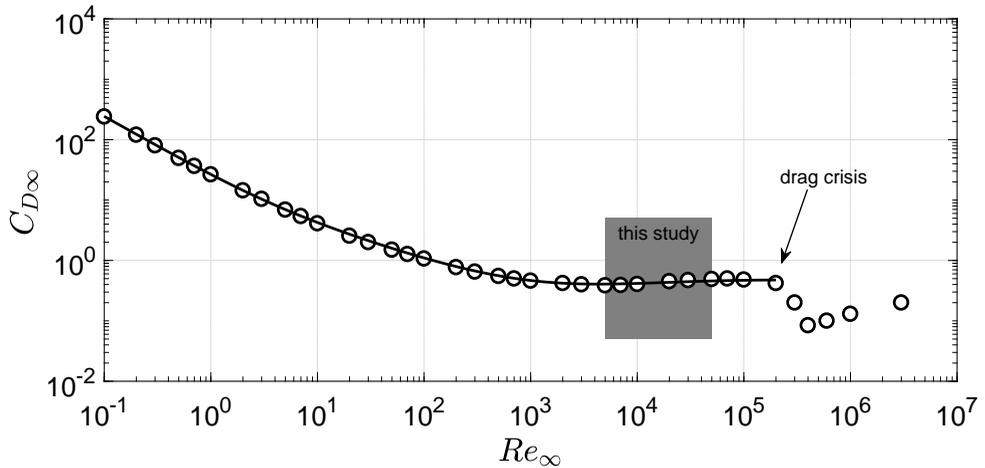}}
  \caption{Drag coefficient as a function of the Reynolds number for smooth spheres. $\circ$: experimental data \citep{Lapple1940}. Solid line: semi-empirical law \citep{Cheng2009} valid up to $\Rey_{\infty} = 2 \times 10^5$. The present study focuses on the sub-critical regime, before the appearance of the drag crisis phenomenon.}
\label{fig:1}
\end{figure}

This study aims at providing new insights about the physical mechanisms governing the development of the wake of SH spheres. To the best of our knowledge, no attempt has yet been undertaken to study the influence of the plastron compliance. This is the main goal of this work. The motivation of this study is twofold: \emph{i.} evidencing and characterising the interface distortion and \emph{ii.} investigating the feedback of the plastron deformation onto the wake of the sphere. To this end, a particular attention is given to the onset of the wake instabilities and the hydrodynamic loads experienced by the SH spheres. In order to highlight the role of the plastron deformation, the dependency of the flow on other parameters must be as small as possible. In particular, the flow regime must be chosen to be Reynolds independent. The well known evolution of the terminal drag coefficient $C_{D \infty}$ of a sphere is plotted in figure \ref{fig:1} with respect to $\Rey_\infty$ \citep{Lapple1940}. Within this study, the focus is put on the so-called sub-critical regime, which is featured by the massive separation of the laminar boundary layer \citep{Achenbach1972}. As illustrated in figure \ref{fig:1}, the drag coefficient characterising this regime is approximately constant, which means that the flow is weakly dependent on Reynolds number. 

The paper is organized as follows: the experimental set-up and the manufacturing procedure of the SH coatings are described in \S\ref{sec:2}. The deformation of the plastron is investigated in \S\ref{sec:3}, while its influence on the hydrodynamic performances of the falling spheres is discussed in \S\ref{sec:4}.

\section{Experimental set-up and manufacturing procedure}
\label{sec:2}
\subsection{The falling sphere experiment}
The experiments were performed in a transparent tank ($100\times100$ mm$^{2}$ cross-section and 650 mm height) filled with double-distilled water (see figures 2(a) and 2(b)). In this study, stainless steel bearing spheres with nominal diameter $d$ ranging from 5 mm to 25 mm were used as reference. Before each try, the spheres were gently dived beneath the water surface by means of an electromagnetic holder, which was used to accurately control their release. Blank tests were performed in order to exclude significant effects of the electromagnetic field on the sphere drops. During the tests the temperature of the water was monitored by a thermocouple and was kept around 19 $^{\circ}$C.

\begin{figure}
  \centerline{\includegraphics[width = 0.95 \textwidth]{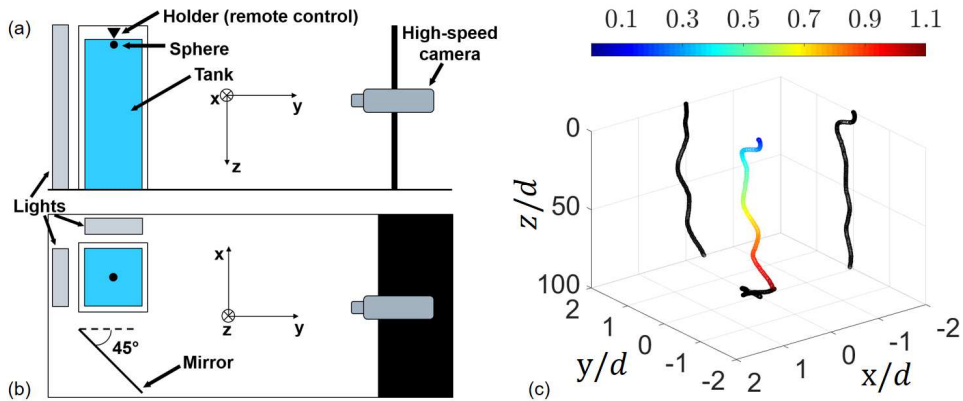}}
  \caption{The falling sphere experimental set-up: (a) side view and (b) top view. (c) Typical example of a 3D trajectory of a free falling sphere. The color scale indicates the sphere velocity in ms$^{-1}$. The black curves represent the projections of the trajectory on the three planes.}
\label{fig:2}
\end{figure}

Once released, the sphere motion was recorded by a Phantom V341 high-speed camera (Zeiss Makro-Planar T* 2/50 mm ZF lens), at a $2560\times1100$ px$^{2}$ resolution with a frame rate of 1000 fps. A mirror placed at 45$^{\circ}$ with respect to the tank enabled the simultaneous recording of front and side views with a single camera. The selected recording window resulted into a conversion factor of 0.3 mm/px. The results reported in this paper are obtained by averaging ten independent tries. A settling time was imposed between each test to ensure the water to be completely at rest before the sphere release. A post-processing analysis of the recorded videos, carried out with the commercial software MATLAB\textsuperscript{\textregistered}, allowed the reconstruction of the time-resolved 3D trajectory of the falling spheres, with the $z$-axis aligned with the gravity $g$ and the $x$ and $y$ coordinates referring to the horizontal plane. The evolution of the sphere velocity was deduced from the time derivative of these trajectories (see figure 2(c)), with uncertainties lower than 5$\%$ in all cases.

In addition to the sphere motion measurement, high magnification videos were performed to investigate the behaviour of the air plastron during the fall. In that case, a $2048\times1152$ px$^{2}$ recording window (Sigma 180 mm F2.8 APO Macro EX DG OS lens) and a 1300 fps recording speed were used. Only the front plane was recorded, and the corresponding spatial resolution was 0.06 mm/px. 

\subsection{The super-hydrophobic coatings}
A wide range of manufacturing procedures have been developed to produce SH surfaces \citep[see e.g.][for a review]{Zhang2008, Bhushan2011}, from chemical vapour deposition \citep{Lau2003} to costless mechanical sanding of naturally water-repellent materials \citep{Nilsson2010} to cite but a few. Depending of the selected technique, either regular or random patterns of micron-sized and/or nano-sized structures can be produced at the surface. In most cases, the samples studied are flat surfaces. However, a large number of the techniques mentioned above are not suited as far as curved surfaces are concerned. This is the reason why a spray coating method \citep{McHale2009, Aljallis2013, Kim2015} was used to produce the SH spheres. The SH properties were obtained by depositing a commercially available product, Ultra-Ever Dry\textsuperscript{\textregistered} \citep{UED}. To this end, a specific manufacturing procedure was designed as follows:
\begin{enumerate}
\item the surface of the reference sphere was first cleaned with acetone then rinsed with double-distilled water and finally dried (see figure 3(a)),
\item an etch primer (Ultra-Ever Dry\textsuperscript{\textregistered} bottom coating) was uniformly sprayed over the surface,
\item an intermediate layer made of a carbon-based powder was deposited onto the surface in order to control the surface texture (see figure 3(b)), \label{step:3}
\item the etch primer was sprayed again over the intermediate layer and then air dried for at least 1 hour with a double purpose: firmly stick the powder to the surface and provide a consistent material for the SH coating to bond, \label{step:4}
\item the SH properties were obtained by spraying the Ultra-Ever Dry\textsuperscript{\textregistered} top coating over the surface and letting it dry for at least 2 hours.
\end{enumerate}

Two different carbon-based powders with grade P220 and P80 were employed as intermediate layers in this study. The resulting coatings are labelled SH-220 and SH-80, respectively. In addition, SH spheres without intermediate layer were produced by skipping steps 3 and 4 of the described procedure. This coating will be referred to as SH-NAR (No Additional Roughness) in the following. Compared to the smooth spheres, the deposit of the SH coating increases both mass and diameter, which have been measured with an accuracy of 0.1 mg and 10 $\mbox{\SI{}{\micro\metre}}$, respectively. Accordingly, the resulting density ratio between sphere and water ($\zeta=\rho_{s}/\rho_{f}$), which is one of the main driving parameter of the falling motion \citep{Ern2012} lays within [6.2 - 7.8]. The high values of $\zeta$ achieved in this study imply that the sphere motion is predominantly vertical (see figure 6 in \citet{Ern2012} where this behaviour is referred to as mode A) as illustrated by the scales of movement in figure 2(c). Note also that the size of the roughness deposited on the sphere surface is not high enough to promote the laminar-to-turbulent transition of the boundary layer \citep{Achenbach1974} which would lead to the so-called drag crisis (see figure \ref{fig:1}).

In addition, SH flat plates were produced following the same manufacturing procedure in order to analyse the properties of the SH coatings. Results obtained from the combination of confocal microscopy (see figure \ref{fig:4}) and contact angle measurements are outlined in table \ref{tab:1}, which emphasises the change in surface roughness $\lambda$ as a function of the grade of the intermediate layer. All produced surfaces are characterised by high static contact angle $\theta _{s}$ and low roll-off angle $\theta _{r}$ (indicative of small contact angle hysteresis), which are symptomatic of the so-called Cassie-Baxter state. This is well supported by the presence of the air plastron around SH spheres once immersed in water as demonstrated in figure 3(c).

\begin{figure}
  \centerline{\includegraphics[width = 0.50 \textwidth]{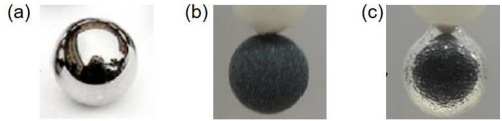}}
  \caption{Images taken during the manufacturing procedure of a $d=10$ mm SH sphere illustrating: (a) reference stainless steel sphere, (b) intermediate layer made of a carbon-based powder, (c) air plastron around the surface of the SH sphere once immersed in water.}
\label{fig:3}
\end{figure}

\begin{table}
\begin{center}
\def~{\hphantom{0}}
\begin{tabular}{lccc}
 & SH-NAR & SH-220 & SH-80\\
$\lambda$ [$\mbox{\SI{}{\micro\metre}}$] & 25 & 74 & 142\\
$\theta _{s}$ [$^{\circ}$] & $160.7 \pm 1.4$ & $150.1 \pm 1.5$ & $145.7 \pm 1.0$\\
$\theta _{r}$ [$^{\circ}$] & $1.6 \pm 0.1$ & $2.4 \pm 0.5$ & $5.4 \pm 1.6$\\
\end{tabular}
\caption{Major properties of the produced SH coatings. $\lambda$, root-mean-square surface roughness. $\theta _{s}$, static contact angle. $\theta _{r}$, roll-off angle.}
\label{tab:1}
\end{center}
\end{table}

\begin{figure}
  \centerline{\includegraphics[width = 0.60 \textwidth]{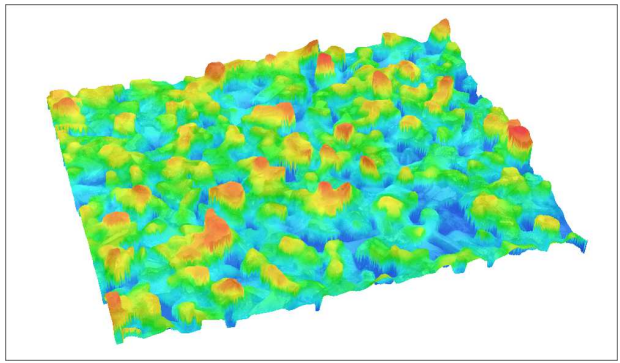}}
  \caption{Confocal microscopy analysis of a portion ($1.5\times1.0$ mm) of a flat plate covered by the SH-220 coating. The color scale (blue-to-red) indicates the surface roughness (0 - 170 $\mu$m). Notice the completely random spatial distribution of the powder particles.}
\label{fig:4}
\end{figure}

\section{Plastron deformation}
\label{sec:3}
For the remainder of the paper, the symbol $^{\star}$ denotes dimensionless variables where the time $t_0=\sqrt{d/\left(\left(\zeta -1\right) g\right)}$ and the velocity $V_0=\sqrt{\left(\zeta -1\right) g d}$ are used as scaling parameters to account for gravity/buoyancy effects \citep{Jenny2004}. Accordingly, the typical Reynolds number is defined as $Re = \rho_f V_0 d / \mu$, with $\mu$ the dynamic viscosity of water. Note that in this study $\zeta$ is kept roughly constant, which means that the sphere diameter, $d$, is the main control parameter.

\begin{figure}
  \centerline{\includegraphics[width = 0.95 \textwidth]{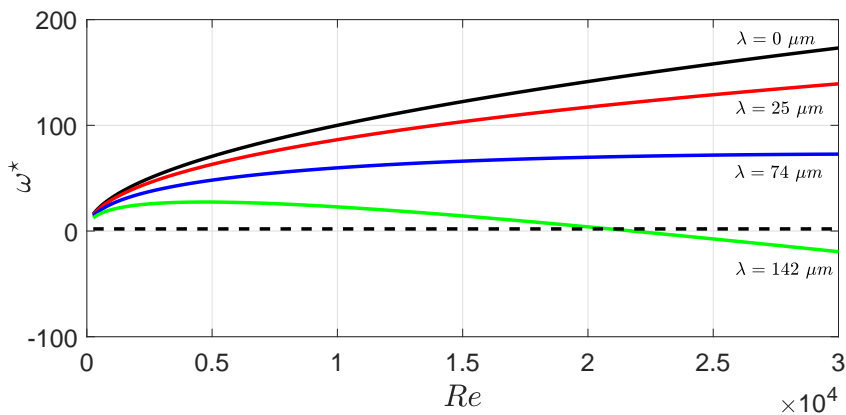}}
  \caption{Predicted vorticity at the surface at a sphere encapsulated by an ideal air layer for different slip length according to equation \ref{eq:vort}. Solid black: $\lambda = 0$ \SI{}{\micro\meter}. Solid red: $\lambda = 25$ \SI{}{\micro\meter}. Solid blue: $\lambda = 74$ \SI{}{\micro\meter}. Solid green: $\lambda = 142$ \SI{}{\micro\meter}. The dashed line represents the limit beyond which the vortex shedding is likely to appear \citep[see e.g.][]{Mougin2002b, Legendre2009}.}
\label{fig:5}
\end{figure}

As mentioned previously, the amount of vorticity generated at the surface of the body is a key ingredient for what concerns the development of the wake. The production of vorticity at the sphere surface is expected to be concentrated within the laminar boundary layer, which is characterised by its thickness $\delta \sim d Re^{-1/2}$. The vorticity generated in the boundary layer can be approximated by $\omega \sim \Delta U / \delta$ with $\Delta U$ the typical velocity jump across the boundary layer \citep{Batchelor1967}. In the case of a slippery wall, $\Delta U \sim V_0 - V_g$ where $V_g$ is the slip velocity. Introducing the slip length model and using the surface roughness $\lambda$ as slip length \citep{Ybert2007} yields $V_g \sim \lambda u_\tau^2 / \nu$, with $u_\tau$ the friction velocity. Assuming the viscous drag to account for 10\% of the total drag (sub-critical regime) we obtain $u_\tau = k V_0$, with $k =$ \textit{O}$\left(10^{-1}\right)$. Plugging these expressions together yields:

\begin{equation}
\omega^\star = A \left(1 - k \frac{\lambda}{\delta_\nu}\right) \frac{d}{\delta},
\label{eq:vort}
\end{equation}

with $\delta_\nu = \nu/u_\tau$ the viscous length scale and $A =$ \textit{O}$\left(1\right)$. For a given sphere diameter, this relationship predicts that the production of vorticity decreases linearly with the slip length, which is assimilated to the surface roughness here. The evolution of the dimensionless vorticity computed according to equation  \ref{eq:vort} is plotted in figure \ref{fig:5} for the different surface coatings used in this study. While each SH coating leads to a reduction of the vorticity, values lower than the threshold at which vortex shedding is expected to occur \citep{Mougin2002b, Legendre2009} are predicted for the SH-80 coating at large $Re$. However, it is important to recall that these predictions are based on the assumption that the air layer encapsulating the sphere is non-deformable. This hypothesis is assessed in the following.

\begin{figure}
  \centerline{\includegraphics[width = 0.95 \textwidth]{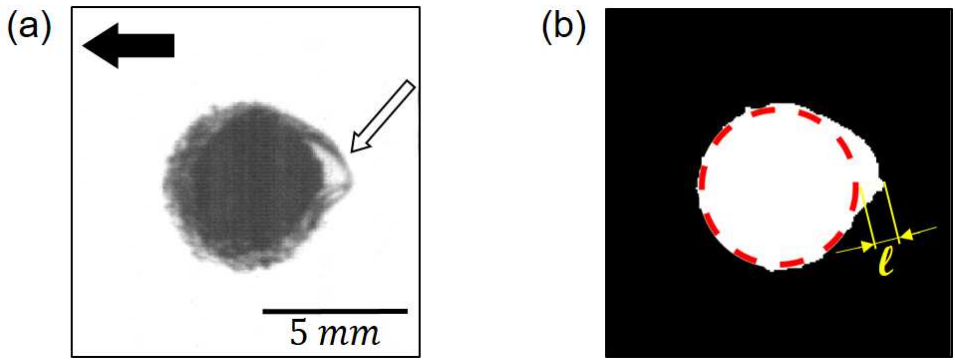}}
  \caption{(a) Typical image recorded during the fall of a SH-80 sphere ($d= 5$ mm) using the high magnification configuration. The black arrow denotes the gravity direction. The white arrow indicates the region where the plastron deformation is visible. (b) Binarized image resulting from the contour finding algorithm used to compute the aspect ratio $\chi$. The red dashed line symbolises the sphere. The variable $\ell$ denotes the typical length scale of the observed protrusion.}
\label{fig:6}
\end{figure}

\begin{figure}
  \centerline{\includegraphics[width = 0.95 \textwidth]{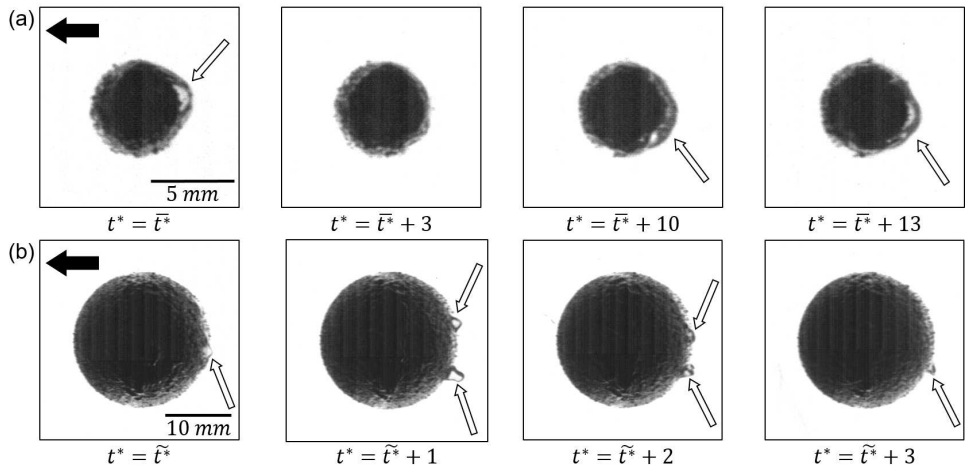}}
  \caption{Typical snapshots recorded in high magnification configuration illustrating the movement and deformation of the air plastron (indicated by white arrows) around SH-80 coated spheres. (a) $d=5$ mm, (b) $d=20$ mm. The black arrows represent the gravity direction. The variables $\bar{t^\star}$ and $\tilde{t^\star}$ designate two time origins chosen randomly.}
\label{fig:7}
\end{figure}

Figure \ref{fig:6} shows a typical snapshot of a SH-80 sphere ($d = 5$ mm) during its falling motion using the high magnification configuration. This image clearly illustrates the deformation of the air plastron as testified by the presence of a huge protrusion located at the backside of the sphere. The plastron deformation can be quantified by introducing the aspect ratio $\chi$ such that:

\begin{equation}
\chi = \frac{d_{eq}}{d},
\label{eq:aspectratio}
\end{equation} 

where $d_{eq}$ denotes an equivalent diameter defined as follows:

\begin{equation}
d_{eq} =  2\sqrt{\frac{\mathcal{S}}{\pi}},
\end{equation}

with $\mathcal{S}$ the surface area delineated by the deformed interface. Based on the MATLAB\textsuperscript{\textregistered} image processing toolbox, a contour finding algorithm has been developed to estimate $\mathcal{S}$. The output of this algorithm applied to the image shown in figure \ref{fig:6}(a) is illustrated in figure \ref{fig:6}(b). In practice, $\mathcal{S}$ is estimated via the computation of the white area. As shown in figure \ref{fig:7}, the plastron shape is extremely sensitive to the sphere diameter or equivalently $Re$. Indeed, while the sequence in figure \ref{fig:7}(a) ($d = 5$ mm) is characterized by the presence of a single protrusion, the sequence in figure \ref{fig:7}(b) ($d = 20$ mm) reveals the presence of multiple air pockets. Note however that the typical length scale, $\ell$, of the protrusions observed in this study (see figure \ref{fig:6}(b)) remains roughly constant ($\approx 1.5 \pm 0.5$ mm) independently of $d$ and $\lambda$. From the image post-processing of the high magnification configuration, it is found that $\chi$ decreases from around 1.2 for the smallest SH spheres (i.e., $d = 5$ mm) to about 1 for the largest SH spheres (i.e., $d = 25$ mm). In other words, although the local normalized curvature $d/\ell$ increases with the sphere diameter, at the same time the global shape of the plastron tends to be spherical. This means that the idealized shape assumed to derive the expression \ref{eq:vort} is more likely to be achieved for the largest sphere diameter in this study.

As emphasised in figures \ref{fig:6} and \ref{fig:7}, the interface distortion occurs at the backside of the spheres, which undergoes the massive separation of the laminar boundary layers typical of the sub-critical regime. This suggests that the pressure deficit induced by the flow separation promotes the (partial) suction of the air entrapped in the surface roughness. The change in $\chi$ is therefore the result of the local competition between the separation-induced suction ($\sim \rho V_0^2$) and the capillary pressure ($\sim \gamma /d$, with $\gamma$ the water/air surface tension), which can be qualitatively represented by introducing the Weber number such as:

\begin{equation}
\mathcal{W}e = \frac{\rho_f V_0^2 d}{\gamma}.
\end{equation}

It is worth noticing that this definition differs from that proposed by \cite{Seo2015} who quantified the flow/interface interaction in a turbulent channel flow with viscosity-based variables. Indeed, the wake of the sphere studied here is predominated by the form drag. However, one can show that $\mathcal{W}e^+ = \mathcal{W}e k^2 \delta_\nu /d$, where $\mathcal{W}e^+$ is the Weber number in wall units as introduced by \cite{Seo2015}. In this study, $\mathcal{W}e^+ \approx 10^{-3}$. Furthermore, it is found that $\lambda^+ = \lambda u_\tau / \nu \leq 15$ which is lower than the critical slip length where plastron failure is expected to occur as predicted by \citet{Seo2015}. This is consistent with the fact that we did not observe the release of air bubbles in the wake except when the sphere hit the bottom of the tank.

According to the definition of $V_0$, it comes:

\begin{equation}
\mathcal{W}e = \frac{\rho_f g}{\gamma} \left(\zeta - 1\right) d^2.
\label{eq:3.4}
\end{equation}

Recalling that here $\zeta$ is roughly constant, this expression implies that the sphere diameter is also the main control parameter of the flow/interface interaction.

An important remark has to be made here about the relationship between $\chi$ and $\mathcal{W}e$. Although both of them characterise the interaction between the plastron and the local stresses induced by the wake, $\chi$ is perfectly suited to deliver an \emph{a posteriori} quantitative description of the interface deformation. However, a reliable estimation of $\chi$ suffers from a severe limitation. Indeed, the computation of $\chi$ is based on a two-dimensional projection of a phenomenon which is likely three-dimensional. Unfortunately, the out-of-plane deformations, although perceptible (see supplementary material), are hidden due to the sphere opacity. This is the reason why in this study $\mathcal{W}e$ is more convenient to be used rather than $\chi$. In fact, since $\mathcal{W}e$ is set \emph{a priori}, it should be regarded as a qualitative indicator of plastron deformation. 
Let us now derive a relation linking $\chi$ and $\mathcal{W}e$. Assuming that $d_{eq} \approx d + \ell$, equation \ref{eq:aspectratio} can be reformulated as follows:

\begin{equation}
\chi = 1 + \frac{\ell}{d}.
\label{eq:csi}
\end{equation}

Since $\ell$ is found almost constant, it comes:

\begin{equation}
\mathcal{W}e \sim \left(\chi - 1\right)^{-2}.
\label{eq:We_csi}
\end{equation}

In other words, low values of $\mathcal{W}e$ are associated to large $\chi$ and \emph{vice versa}. The implications of this relationship will be discussed further in \S\ref{sec:4.3}.\\

\begin{figure}
  \centerline{\includegraphics[width = 0.6 \textwidth]{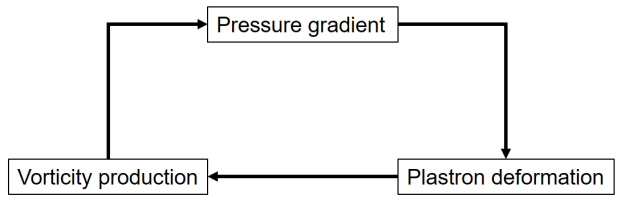}}
  \caption{A possible driving mechanism of the mutual interaction between the flow and the air plastron.}
\label{fig:8}
\end{figure}

The visualisations provided in figures \ref{fig:6} and \ref{fig:7} suggest that the air plastron compliance and its dynamics are intimately connected to the wake. This raises the issue of a possible feedback of the plastron motion and deformation onto the flow. Somehow, the plastron deformation evidenced in this study shares some properties of freely rising bubbles \citep{Ellingsen2001}. Basically, the main difference is that bubbles adapts their entire shape whereas here only a thin layer is compliant. It is the reason why the maximum values of $\chi$ observed here ($\approx 1.2$) are much lower compared to those of fully deformable bodies \citep{Mougin2002}. Nevertheless, it is reasonable to assume that the physical mechanisms driving the body/flow interaction are comparable, at least qualitatively. In figure \ref{fig:8}, a schematic description of this interaction is proposed: the unsteady pressure gradient induced by the massive separation around the falling sphere promotes the deformation and the movement of the air plastron. This modification of the boundary conditions may alter significantly the production of vorticity at the body surface and accordingly the intensity of vorticity within the wake \citep[see e.g.][]{Mougin2002,Legendre2009}, which in turn may impact the pressure gradient. The feedback loop described in figure \ref{fig:8} suggests that the plastron deformation could have a sizeable effect on the sphere wake since the hydrodynamic forces experienced by the body are intimately linked to both pressure and vorticity distributions. 

\section{Hydrodynamic performances}
\label{sec:4}
So far, our observations reveal that the plastron shape is extremely sensitive to the stresses induced by the wake. To proceed further, we now investigate the influence of physical parameters on the hydrodynamic forces, i.e. lift and drag, experienced by the falling spheres, with the goal to emphasise the feedback of plastron deformation onto the flow. The effects of three possible parameters could be assessed: \emph{i.} $Re$ which characterizes the wake, \emph{ii.} $\lambda/d$ which is representative of the slip and 
\emph{iii.} $\mathcal{W}e$ (or equivalently $\chi$) which describes the interface deformation. Note that due to their definitions, $Re$ and $\mathcal{W}e$ cannot be set independently. However, as mentioned previously, we expect the physical influence of $Re$ to be negligible (at least over the range studied here) in comparison of that of $\mathcal{W}e$. This point will be discussed further in \S\ref{sec:4.3}. Based on this assumption, we focus mostly on the role of $\lambda/d$ and $\mathcal{W}e$ in the following.

\subsection{Vertical motion}
\label{sec:4.1}
\begin{figure}
\centerline{\includegraphics[width = 0.86 \textwidth]{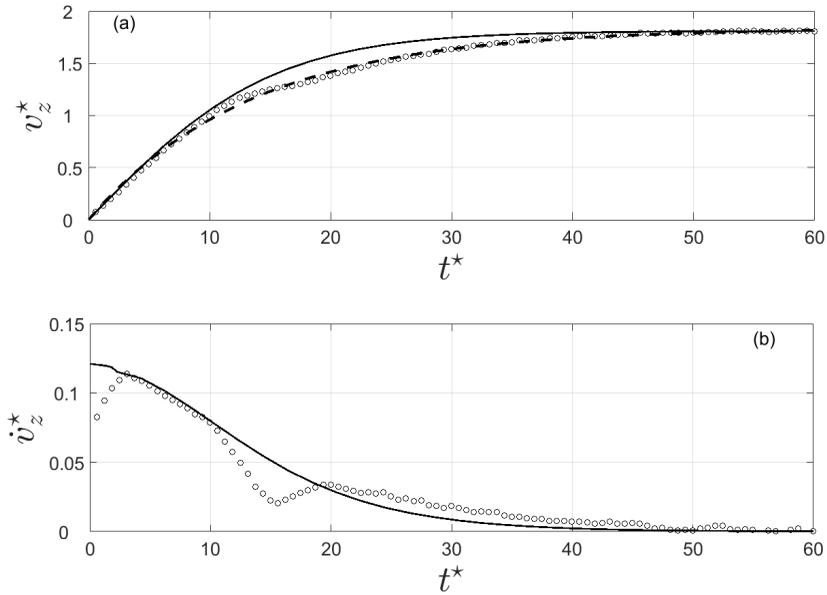}}
\caption{Time evolution of the vertical velocity (a) and acceleration (b) of the $d=5$ mm reference (smooth) spheres. $\circ$: experimental data (for the sake of readability, 1 point out of 2 is reported). Solid line: prediction according to equation \ref{eq:1}. Dashed line: fitted model according to equation \ref{eq:2} \citep{Mordant2000}.}
\label{fig:9}
\end{figure}

\begin{figure}
\centerline{\includegraphics[width = 0.86 \textwidth]{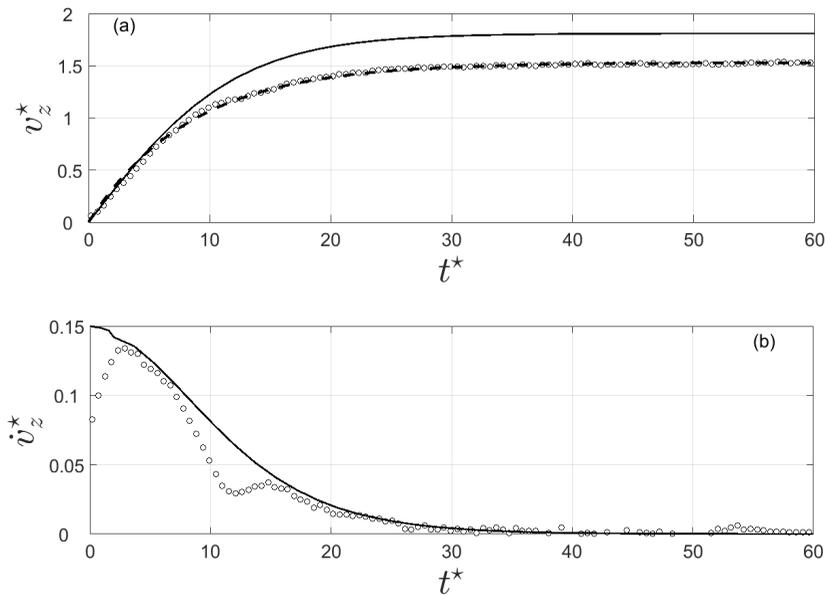}}
\caption{Time evolution of the vertical velocity (a) and acceleration (b) of the $d=5$ mm SH-80 coated spheres. $\circ$: experimental data (for the sake of readability, 1 point out of 2 is reported). Solid line: prediction according to equation \ref{eq:1}. Dashed line: fitted model according to equation \ref{eq:2} \citep{Mordant2000}.}
\label{fig:10}
\end{figure}

Let us start by investigating the fall of a sphere for which large deformations happen, i.e. at low $\mathcal{W}e$. The time histories of the normalised vertical falling velocity $v_z^{\star}$ and acceleration $\dot{v}_{z}^{\star}=\frac{dv_z^\star}{dt^\star}$ for the smallest spheres ($d = 5$ mm) are displayed in figures \ref{fig:9} and \ref{fig:10}, comparing the reference and the SH-80 coated surfaces. Starting from rest the sphere first accelerates yielding an increase of $v_z^{\star}$ and then reaches its terminal velocity $v_{z\infty}^{\star}$ once transient effects vanish. Note that, in this study, since $\zeta \approx 7$, the vertical motion could be well approximated by solving the following equation \citep{Mordant2000}:
\begin{equation}
\left(\zeta + \frac{1}{2}\right) \dot{v}_z^\star = 1 - \frac{3 C_D}{4}  v_z^{\star 2},
\label{eq:1}
\end{equation}
where the empirical law $C_{D}(\Rey)$ given by \citet{Cheng2009} was considered (see figure \ref{fig:1}). In equation \ref{eq:1}, the left-hand side term is the acceleration weighted by the added mass, while the right-hand side terms represent the gravity/buoyancy and the drag force respectively. For the sake of simplicity, the term accounting for the Stokes memory effects was not considered (see \citet{Mordant2000} for full details). Note that at release time, i.e. $t^{\star} = 0$, the drag force is null. This, combined with the exclusion of memory effects implies that the predicted velocity is expected to be slightly overestimated at the very beginning of the sphere motion.

For the reference sphere, a good agreement between the predicted and the measured falling velocity is observed at the early stage of the motion and in the steady state as well (see figure \ref{fig:9}(a)). However, a significant departure arises within the range $t^{\star} \in \left[10; 40\right]$. The experimental velocity drop is induced by an acceleration decrease occurring between $t^{\star} \approx 10$ and $t^{\star} \approx 20$ (see figure \ref{fig:9}(b)). Such phenomenon has been reported by other authors and is induced by the onset of instabilities in the wake of the sphere, which promotes a sudden drag increase \citep{Jenny2004}. This point will be discussed further in \S\ref{sec:4.2}. More surprisingly, from $t^{\star} \approx 20$ to $t^{\star} \approx 40$ the measured sphere acceleration exceeds its predicted value, which balances the previous acceleration reduction and enables the reference sphere to reach its predicted terminal velocity, provided that $v_{z\infty}^{\star} = \int_{0}^{\infty} \dot{v}_{z}^{\star} dt^{\star}$.

For the SH sphere, figure \ref{fig:10}(b) similarly evidences an acceleration decrease. However, the departure from the predicted acceleration occurs at $t^{\star} \approx 7$, which means that the SH coating promotes the onset of instabilities earlier than the reference sphere. Moreover, unlike the reference sphere, the measured acceleration then collapses on the predicted trend beyond $t^{\star} \approx 15$. Accordingly, the terminal velocity achieved by the SH sphere is much lower than its predicted value as emphasised in figure \ref{fig:10}(a). In the specific case displayed in figures \ref{fig:9} and \ref{fig:10}, i.e. $d = 5$ mm, this yields an increase of terminal drag coefficient due to the SH coating, since equation \ref{eq:1} reduces to $C_{D \infty} = \frac{4}{3 v_{z \infty}^{\star 2}}$ once transient effects vanish. This result agrees with the findings of \citet{Ahmmed2016} who reported drag increase of laser-textured SH spheres at comparable Reynolds numbers.

\subsection{Lift force}
\label{sec:4.2}
\begin{figure}
\centerline{\includegraphics[width = 0.95 \textwidth]{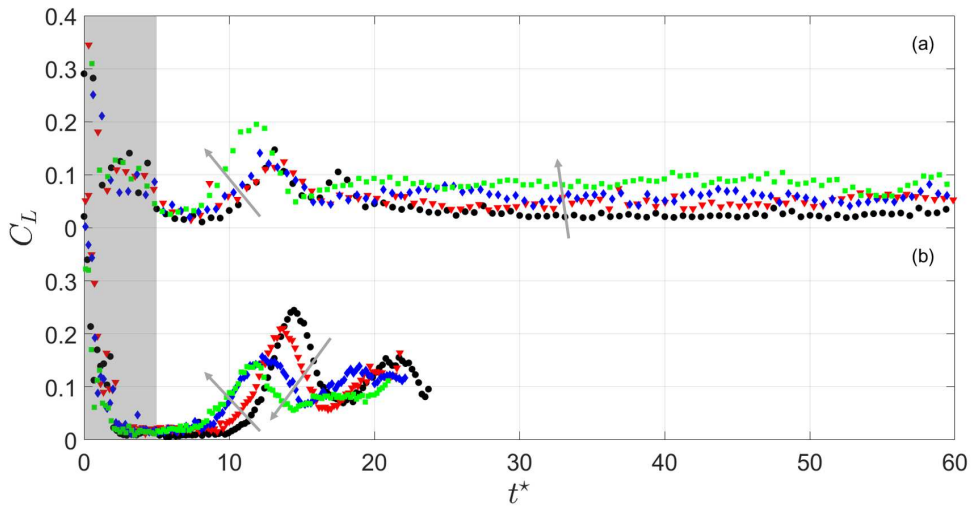}}
\caption{Time history of the lift coefficient of the (a) $d = 5$ mm and (b) $d = 20$ mm spheres (for the sake of readability, 1 point out of 2 is reported). $\bullet$: reference (smooth) sphere. $\textcolor{red}{\blacktriangledown}$, SH-NAR coating. $\textcolor{blue}{\blacklozenge}$, SH-220 coating. $\textcolor{green}{\blacksquare}$, SH-80 coating. The grey arrows indicate increasing roughness thickness. The shaded region identifies the portion of time where lift estimation is corrupted by the signal-to-noise ratio.}
\label{fig:11}
\end{figure}

In what follows we focus on the flow-induced loads experienced by the spheres, starting with lift. Since lift force results from the asymmetric distribution of vorticity released in the wake, it is a very well suited indicator of instability onset. The time history of the lift coefficient $C_{L} = 8 \vert L \vert /\left(\rho_{f} v_{z}^{2} \pi d^{2}\right)$, where $\vert L \vert$ is the lift force magnitude, was estimated by solving the generalized Kirchhoff equations \citep{Mougin2002b} following the methodology described in detail by \citet{Shew2006}. Figure \ref{fig:11} shows the evolution of $C_{L}$ calculated from our experimental data for $d = 5$ mm ($\mathcal{W}e \approx 20$) and $d = 20$ mm ($\mathcal{W}e \approx 360$) spheres. Note that the signal-to-noise ratio is too high at the early stage of motion (shaded zone) for the lift force estimation to be reliable. At the starting of motion, $C_{L}$ is almost null, meaning that the wake is axisymmetric and accordingly the trajectory is vertical. Then, $C_L$ suddenly increases as a result of the loss of axisymmetry of the wake, which originates from instabilities arising in the wake \citep{Magnaudet2000,Horowitz2010b}. These instabilities promote path unsteadiness (see figure \ref{fig:2}(c)) together with vertical velocity decrease with respect to the prediction, as discussed in \S\ref{sec:4.1}. However, the amplitude of path excursion is moderate due to lift fluctuation damping caused by the high $\zeta$ values considered in this study. Once lift fluctuations vanish, $C_{L}$ remains almost constant, which implies the sphere to follow an oblique trajectory. Note that for the $d=20$ mm spheres, the tank is not tall enough for the lift force to reach its steady state. Nevertheless, the wake transition as well as the damping of the lift fluctuation are clearly visible.

\begin{figure}
\centerline{\includegraphics[width = 0.95 \textwidth]{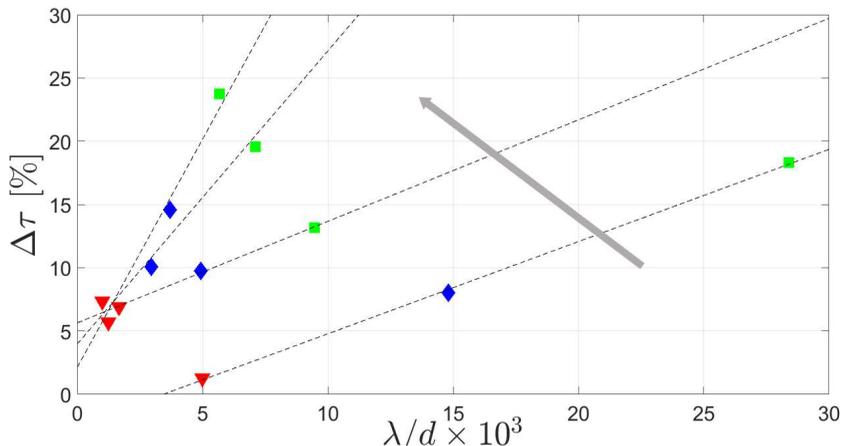}}
\caption{Effect of surface properties on the characteristic time at which wake instabilities appear. $\textcolor{red}{\blacktriangledown}$, SH-NAR coating. $\textcolor{blue}{\blacklozenge}$, SH-220 coating. $\textcolor{green}{\blacksquare}$, SH-80 coating. Dashed lines: best fit of equation \ref{eq:Dt} in the least-mean square sense. The grey arrow indicates increasing $\mathcal{W}e$. For the sake of readability, only results for $d =$ 5, 15, 20 and 25 mm are reported.}
\label{fig:TimeInsta}
\end{figure}

We now introduce $\tau$ as the characteristic time at which the wake instability occurs. In practice, $\tau$ is estimated as the time where the condition $C_L = 0.05$ is first met. The evolution of $\Delta \tau = 1 - \tau^{SH} / \tau^{ref}$ (with superscripts $SH$ and $ref$ referring to SH coating and smooth surface, respectively) is displayed in figure \ref{fig:TimeInsta} with respect to the dimensionless roughness $\lambda/d$ for different sphere diameters. It appears that the data are well fitted by a linear relationship:

\begin{equation}
\Delta \tau \approx \alpha_\tau \frac{\lambda}{d},
\label{eq:Dt}
\end{equation}

with $\alpha_\tau$ the slope of the law. In any case, the SH coatings trigger wake instability earlier than the reference spheres. Our findings agree with the results of \citet{Jenny2004} and \citet{Fernandes2007} who reported earlier transition of asymmetric bodies. In a similar fashion, the imperfect sphericity induced by the interface distortion may trigger the wake instabilities more rapidly. 

\begin{figure}
\centerline{\includegraphics[width = 0.95 \textwidth]{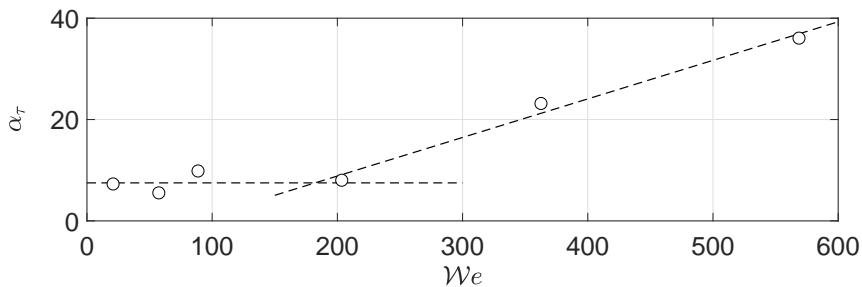}}
\caption{Evolution of the slope $\alpha_\tau$ (see equation \ref{eq:Dt}) with respect to $\mathcal{W}e$. Dashed lines: best fit of equation \ref{eq:alphatau} in the least-mean square sense.}
\label{fig:SlopeTimeInsta}
\end{figure}

Figure \ref{fig:SlopeTimeInsta} shows the variation of $\alpha_\tau$ as a function of $\mathcal{W}e$. It is found that $\alpha_\tau \sim \mathcal{W}e^m$ over the available range of operating conditions tested in this study. Surprisingly, this plot reveals an abrupt change of the power law exponent $m$, which varies from 0 to 1, around a critical value $\mathcal{W}e_c = 180$ such that

\begin{equation}
\left\{
\begin{array}{lclll}
\alpha_\tau & = & \alpha_{\tau 0} & \mbox{if} & \mathcal{W}e < \mathcal{W}e_c \\
\alpha_\tau & = & \alpha_{\tau 0} \left(2\frac{\mathcal{W}e}{\mathcal{W}e_c} - 1\right) & \mbox{if} & \mathcal{W}e \geq \mathcal{W}e_c
\end{array}
\right.
\label{eq:alphatau}
\end{equation}

with $\alpha_{\tau 0} \approx 7.5$. As evidenced in figure \ref{fig:11}, the change in $\alpha_\tau$ observed at $\mathcal{W}e_c$ coincides with a significant modification of the magnitude of both the lift fluctuations during the transient phase and the lift force in the steady regime. For $\mathcal{W}e < \mathcal{W}e_c$ (figure \ref{fig:11}(a)), SH coatings tend to increase lift, especially in the steady state regime. On the contrary, for $\mathcal{W}e \geq \mathcal{W}e_c$ (figure \ref{fig:11}(b)), the lift amplitude is reduced. Let us remind that the lift force is intimately related to the intensity of vorticity around the body. Accordingly, our findings suggest that the high interface distortion which is achieved in average when $\mathcal{W}e < \mathcal{W}e_c$ leads to an increase of the amount of vorticity released in the wake. On the contrary, the production of vorticity seems to be mitigated when the condition $\mathcal{W}e \geq \mathcal{W}e_c$ is met, which would imply that the expression \ref{eq:vort} is assessed.

\subsection{Drag force}
\label{sec:4.3}
\begin{figure}
  \centerline{\includegraphics[width = 0.95 \textwidth]{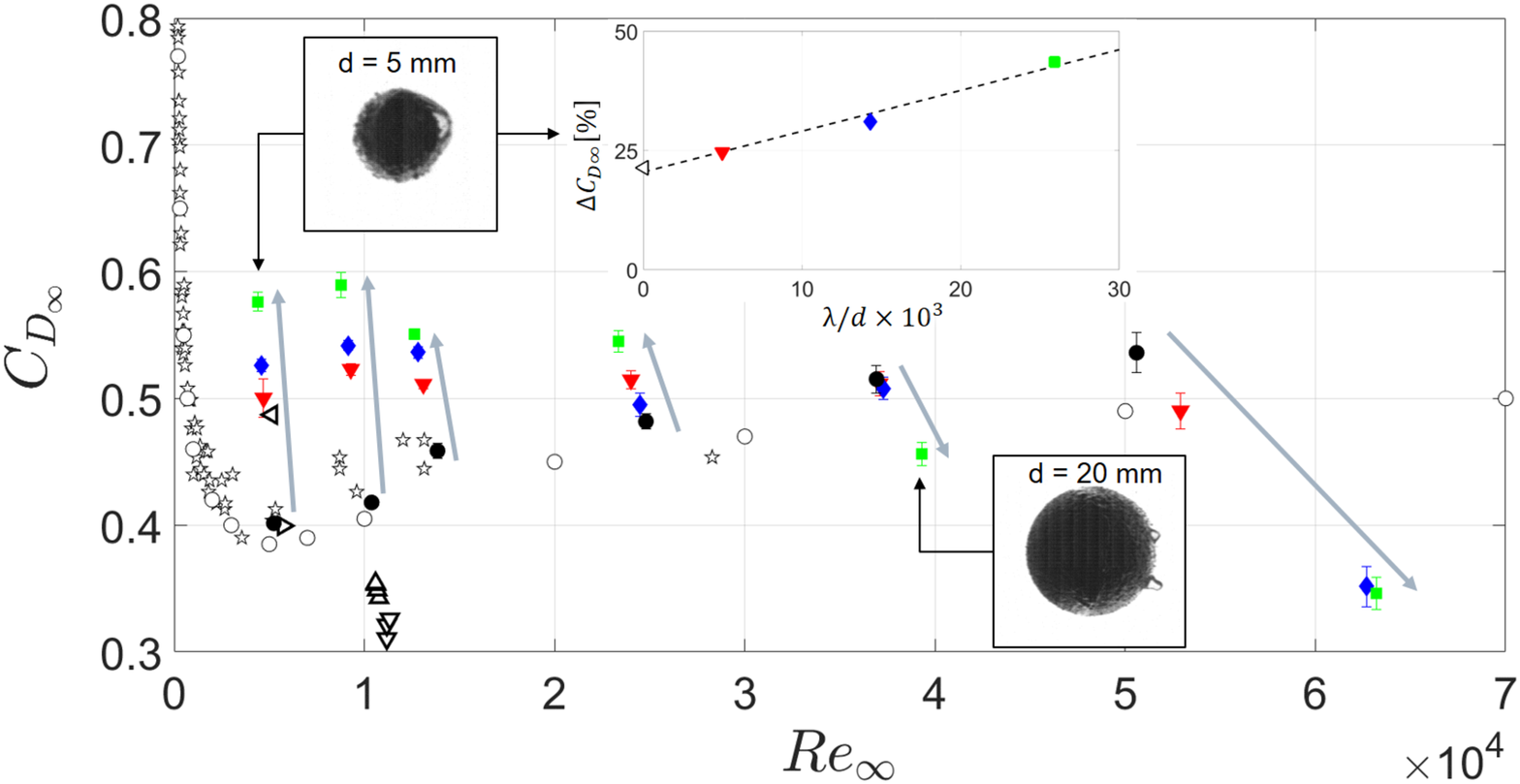}}
  \caption{Terminal drag coefficient $C_{D \infty}$ as a function of terminal Reynolds number $\Rey_{\infty}$ for all the investigated spheres. $\bullet$: reference (smooth) sphere. $\textcolor{red}{\blacktriangledown}$, SH-NAR coating. $\textcolor{blue}{\blacklozenge}$, SH-220 coating. $\textcolor{green}{\blacksquare}$, SH-80 coating. $\star$, spheres falling rectilinearly \citep{Horowitz2010}. $\triangleright$, reference sphere \citep{Ahmmed2016}. $\triangleleft$, SH sphere \citep{Ahmmed2016}. $\bigtriangleup$, $d= 25$ mm reference spheres \citep{McHale2009}. $\bigtriangledown$, $d= 25$ mm SH spheres with air plastron \citep{McHale2009}. $\circ$, experimental data from \citet{Lapple1940}. The grey arrows indicate increasing surface roughness thickness. The inset shows the terminal drag coefficient variation with respect to the reference sphere as a function of the non-dimensional roughness $\lambda/d$ for the spheres in the $\Rey_{\infty} \approx 0.5 \times 10^4$ region.}
\label{fig:14}
\end{figure}

Let us now illustrate the consequences on the terminal drag coefficient $C_{D \infty}$, which is displayed in figure \ref{fig:14} as a function of the terminal Reynolds number $\Rey_{\infty} = v_{z \infty}^\star Re$. As mentioned previously, for the largest spheres ($\Rey_{\infty} \geq 2 \times 10^{4}$) the tank is not tall enough for the steady state to be achieved. In the worst case ($d=25$ mm), the maximum measurable vertical velocity is 15$\%$ lower than the terminal velocity predicted by equation \ref{eq:1}. To overcome this issue, the terminal velocity $v_{z \infty}^{\star}$ was inferred by best fitting, in the least-mean square sense, the data with the following model proposed by \citet{Mordant2000}:
\begin{equation}
v_{z}^{\star} = v_{z \infty}^{\star} \left(1 - e^{-\beta t^{\star}}\right),
\label{eq:2}
\end{equation}
where $v_{z \infty}^{\star}$ and $\beta$ are the fit parameters. Figures \ref{fig:9} and \ref{fig:10} demonstrate the relevance of the model expressed in equation \ref{eq:2}. Anyhow, it is worth noting that due to the limitations of our experimental set-up, the measurements presented in figure \ref{fig:14} for high $\Rey_{\infty}$ should be regarded as qualitative results emphasising the relative influence of plastron properties on drag. A new experimental set-up is currently under design to properly address this issue. For assessment purpose, data obtained by \citet{Lapple1940}, \citet{Horowitz2010} (for spheres falling rectilinearly) as well as those estimated from the results given by \citet{Ahmmed2016} and \citet{McHale2009} for smooth and SH spheres have also been reported. As far as reference (smooth) spheres are concerned, a fairly good agreement between our measurements and the data reported in previous studies is observed over the overlapping $\Rey_{\infty}$ range, with the exception of the work of \citet{McHale2009}. Beyond $\Rey_{\infty} \approx 3 \times 10^{4}$, our data deviates slightly from that of \citet{Lapple1940} which may be due to the fitting procedure.

\begin{figure}
  \centerline{\includegraphics[width = 0.95 \textwidth]{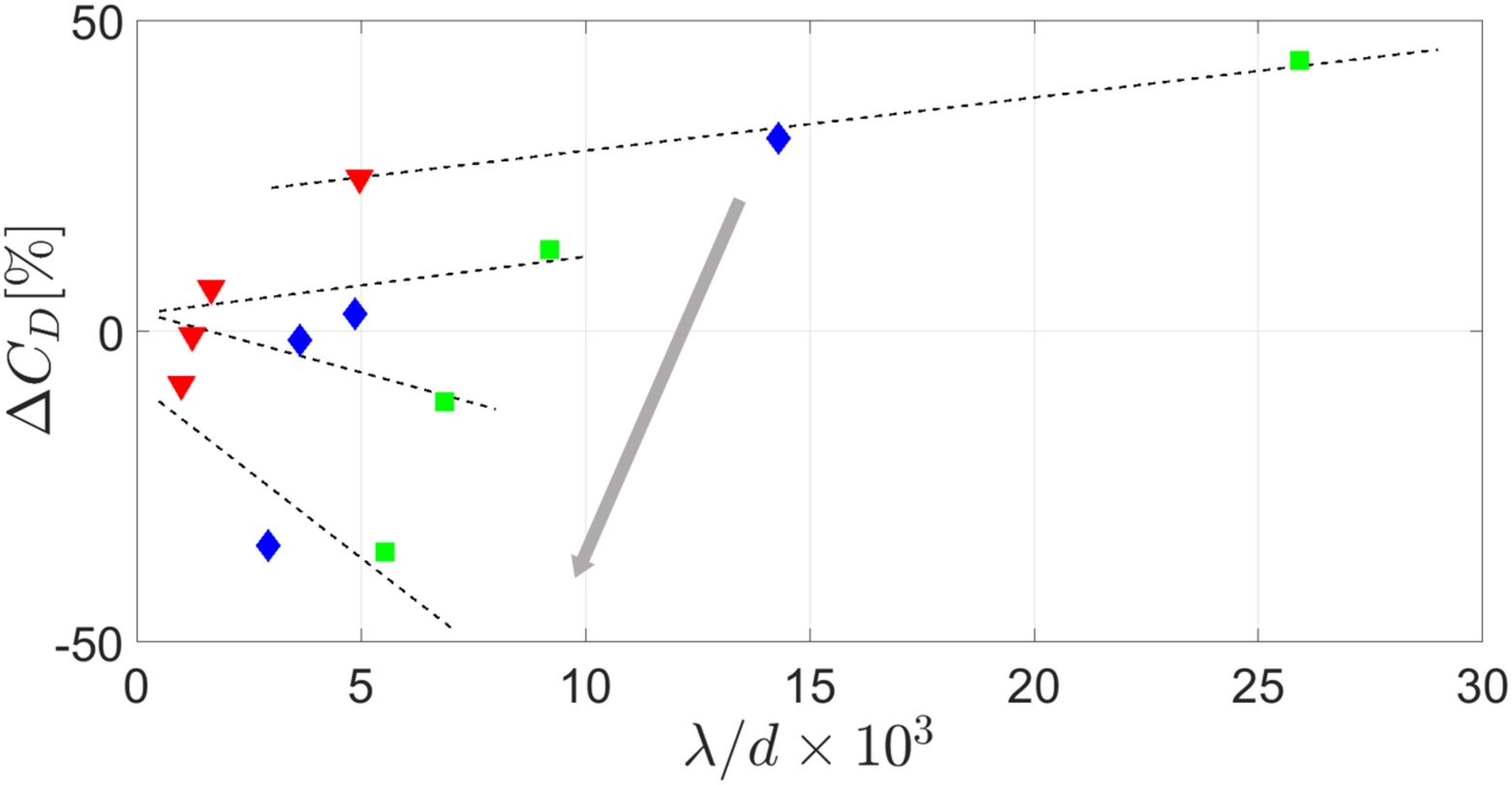}}
  \caption{Terminal drag coefficient variation with respect to the reference sphere as a function of the non-dimensional roughness. $\textcolor{red}{\blacktriangledown}$, SH-NAR coating. $\textcolor{blue}{\blacklozenge}$, SH-220 coating. $\textcolor{green}{\blacksquare}$, SH-80 coating. The dashed lines highlight the linear behaviour (see equation \ref{deltacd}). The grey arrow indicates increasing $\mathcal{W}e$. For the sake of readability, only results for $d =$ 5, 15, 20 and 25 mm are reported.}
\label{fig:15}
\end{figure}

\begin{figure}
  \centerline{\includegraphics[width = 0.95 \textwidth]{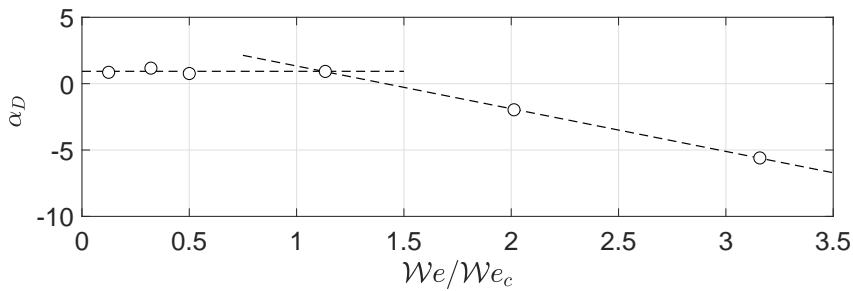}}
  \caption{Slope $\alpha_{D}$ (see equation \ref{deltacd}) as a function of the normalised Weber number for all the tested diameters. $\circ$, experimental data. Dashed lines: best fit of equation \ref{eq:alphaD} in the least-mean square sense.}
\label{fig:16}
\end{figure}

For what concerns SH coatings, both interface deformation (i.e. $\mathcal{W}e$) and slip (i.e. $\lambda/d$) have a sizeable influence on $C_{D \infty}$. We introduce $\Delta C_D = 1 - C_{D \infty}^{SH} / C_{D \infty}^{ref}$ to characterise the drag change induced by the surface properties. The insert in figure \ref{fig:14} shows the evolution of $\Delta C_D$ for the smallest sphere (i.e. $d$ = 5 mm). For comparison, the drag change reported by \citet{Ahmmed2016} at a comparable $Re_\infty$ is also displayed. It is worth noting the excellent agreement between our data and those of \citet{Ahmmed2016}, whose SH spheres are characterised by a nano-sized surface roughness. This plot suggests that the drag change is well approximated by the following law:

\begin{equation}
\Delta C_D = \alpha_D \frac{\lambda}{d} + \Delta C_{D0},
\label{deltacd}
\end{equation}

where $\Delta C_{D0}$ represents the drag change extrapolated at virtually zero surface roughness and $\alpha_D$ stands for the drag rate of change. As evidenced in figure \ref{fig:15}, this trend applies also for the other spheres tested in this study. Drag increase with respect to the reference by up to +45$\%$ is found for the smallest spheres (low $\mathcal{W}e$), whereas a significant drag reduction down to -35$\%$ is noticed for the largest spheres (high $\mathcal{W}e$). This opposite trend is even more evident in figure \ref{fig:16}, which displays the evolution of $\alpha_D$ as a function of $\mathcal{W}e$ normalised by the critical value $\mathcal{W}e_c$.

Pretty much like what was observed for $\alpha_\tau$, it is found that $\alpha_D \sim \mathcal{W}e^m$. Once again, an abrupt change of trend is clearly visible at $\mathcal{W}e_c$ such that:

\begin{equation}
\left\{
\begin{array}{lclll}
\alpha_D & = & \alpha_{D 0} & \mbox{if} & \mathcal{W}e < \mathcal{W}e_c \\
\alpha_D & = & 4.5 - 3.2 \frac{\mathcal{W}e}{\mathcal{W}e_c} & \mbox{if} & \mathcal{W}e \geq \mathcal{W}e_c
\end{array}
\right.
\label{eq:alphaD}
\end{equation}

with $\alpha_{D 0} \approx 0.93$. For $\mathcal{W}e < \mathcal{W}e_c$, $\alpha_D$ is roughly constant and positive meaning that drag increases with increasing plastron thickness. On the contrary, for $\mathcal{W}e < \mathcal{W}e_c$, $\alpha_D$ becomes negative which yields the drag reduction observed in figure \ref{fig:14}.\\

\begin{figure}
  \centerline{\includegraphics[width = 0.95 \textwidth]{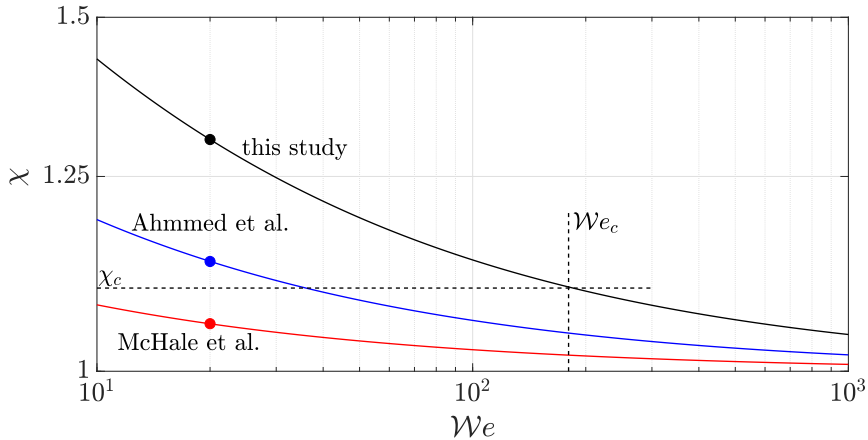}}
  \caption{Variation of the aspect ratio $\chi$ as a function of $\mathcal{W}e$ according to equation \ref{eq:csi_complete}. Solid black: this study. Solid blue: \citet{Ahmmed2016}. Solid red: \citet{McHale2009}. The vertical dashed line indicates the critical Weber number $\mathcal{W}e_{c}$, whilst the horizontal dashed line represents its corresponding critical aspect ratio $\chi_{c}$. The dots symbolise the Weber number region where the studies can be compared ($\mathcal{W}e \approx 20$).}
\label{fig:17}
\end{figure}

It is worth noting that the drag changes observed in this study encompass the different trends reported in previous works \citep{McHale2009, Ahmmed2016}. Taking into account the plastron deformation might provide an attractive way to explain these results within an unified framework. Indeed, according to equation \ref{eq:3.4}, the Weber number characterising the studies of \citet{McHale2009} and \citet{Ahmmed2016} is about 20, which is comparable to the smallest value reached in this work. Based on our results, one would expect drag increase, which is not reported by \citet{McHale2009}. This illustrates that a comprehensive relationship between $\mathcal{W}e$ and $\chi$ needs to be properly established. In fact, a possible explanation of these apparently discordant results arises by developing equation \ref{eq:We_csi} one step further. To this end, equation \ref{eq:3.4} injected into equation \ref{eq:csi} yields:
\begin{equation}
   \chi = 1+\sqrt{\frac{\rho_{f} (\zeta -1) g \ell^{2}}{\gamma}} \mathcal{W}e^{-1/2}.
   \label{eq:csi_complete}
   \end{equation} 
This expression takes into account the influence of $\zeta$, whose values are approximately 2.2 in the work of \citet{Ahmmed2016} (PTFE spheres) and 1.2 in the study of \citet{McHale2009} (acrylic spheres). Based on the assumption that the magnitude of the air layer protrusions, $\ell$, is comparable to that observed here, figure \ref{fig:17} shows the evolution of $\chi$ according to equation \ref{eq:csi_complete}.      
In correspondence to $\mathcal{W}e_c$ (see figure \ref{fig:SlopeTimeInsta}), we introduce a critical aspect ratio $\chi_{c} \approx 1.1$. Based on our findings, highly deformed plastron, i.e. $\chi > \chi_c$, yields drag increase, whilst a plastron almost spherical in average, i.e. $\chi < \chi_c$, results into drag mitigation. Focusing our attention on the $\mathcal{W}e \approx 20$ region, it appears clearly that the aspect ratio featuring the study of \citet{McHale2009} lies below $\chi_c$. This perfectly agrees with the drag reduction these authors reported. Conversely, drag increase observed in this study and in the work of \citet{Ahmmed2016} falls within the region of highly deformed plastron, i.e. $\chi > \chi_c$. All together, our results clearly emphasise that, under the investigated operating conditions, the plastron compliance plays a key role in the performances of SH surfaces. Obviously, further efforts are required to confirm our findings and extrapolate them over a broader range of operating conditions.

\section{Concluding remarks}
In this work, free falling experiments of super-hydrophobic spheres were carried out. To do so, super-hydrophobic coatings were deposited over stainless steel spheres via a spray method suitable for large scale applications. The surface roughness was controlled by embedding micron-sized powders with different particle size during the manufacturing process. A particular attention has been put on the influence of the deformation of the air layer encapsulating the sphere in the Cassie-Baxter state. To this end, the sphere motion was analysed to investigate the hydrodynamic performances in the so-called sub-critical regime.

A first outcome of this study is the evidence of plastron deformation in response to the separation-induced stresses experienced in the wake of the spheres. The competition between the pressure deficit and the surface tension was estimated quantitatively using the aspect ratio $\chi$ of the observed body and qualitatively by introducing the Weber number $\mathcal{W}e$. The analysis of highly resolved visualisations indicates that the plastron adopts an oblate shape at low $\mathcal{W}e$, whilst it tends to be spherical in average at high $\mathcal{W}e$.

The second outcome of this work relates to the feedback of the plastron deformation on the flow. Indeed, it has been found that the plastron compliance has a sizeable influence on the wake development. In a first stage, the transient fall was studied through the time history of the lift coefficient, which is used as a footprint of the amount of vorticity generated during the sphere motion. Compared to smooth spheres used as reference, it has been observed that the onset of the wake instabilities is triggered earlier by the super-hydrophobic coatings. The time at which the instabilities are detected decreases linearly with the slippage effect, which is controlled by changing the surface roughness. Our results highlight an abrupt change of scaling law around a critical Weber number, $\mathcal{W}e_c \approx 200$. A simple model was used to derive the critical aspect ratio $\chi_c$ where the condition $\mathcal{W}e = \mathcal{W}e_{c}$ is met. The steady state is then investigated via the terminal drag coefficient which is compared to values reported in others studies. In comparison to drag measured for the reference spheres, significant drag increase is obtained for highly deformed interface achieved for $\chi > \chi_c$, whereas low deformation obtained for $\chi < \chi_c$ results in drag mitigation. 

Even though the proposed scenario provides an attractive way to explain and unify apparently dissonant results reported in previous works, it requires additional studies with the aim to deliver more reliable SH surface models. Moreover, our study should be extended to a broader range of operating conditions. The motivations of doing so are threefold: \emph{i.} delineate regimes where the plastron compliance cannot be neglected, \emph{ii.} discriminate the effect of $\Rey_\infty$ and $\mathcal{W}e$ and \emph{iii.} assess the relationship between $\mathcal{W}e$ and the aspect ratio $\chi$ outside the sub-critical regime. All together, our findings suggest that the plastron compliance may be a fundamental phenomenon involved in the production of vorticity at the body surface, which in turns deeply impacts the wake. A full understanding of the physical mechanisms at play during the wake transition and the terminal state would require further investigations. Finally, it should be determined whether or not our findings can be extrapolated to other flows (e.g. pressure-driven flows) over SH surfaces.\\

This work was supported by the Direction G\'en\'erale de l'Armement (DGA), Minist\`ere de la D\'efense, R\'epublique Fran\c caise and the Agence Nationale de la Recherche (ANR) through the Investissements d'Avenir program under the Labex CAPRYSSES Project (ANR-11-LABX-0006-01). The financial support of the R\'{e}gion Centre/Val de Loire through the regional project Modif'Surf is also gratefully acknowledged. NM wishes to thank Prof. A. Bottaro and Dr. R. Garc\'{i}a-Mayoral for fruitful discussions.

\bibliographystyle{jfm}
\bibliography{Biblio}

\end{document}